\documentclass[11pt]{article}

\usepackage[margin=1in]{geometry}
\usepackage{amsmath,amssymb,amsfonts,amsthm}
\usepackage{mathtools}
\usepackage{booktabs}
\usepackage{enumitem}
\usepackage{hyperref}
\usepackage{graphicx}
\usepackage{subcaption}
\usepackage{algorithm}
\usepackage{algorithmic}
\usepackage{natbib}
\usepackage{placeins}

\graphicspath{{figures/}{marketgan_binary_medium_constant_outputs/}{marketgan_binary_baseline_constant_outputs/}}

\hypersetup{colorlinks=true,linkcolor=blue,citecolor=blue,urlcolor=blue}

\title{Endogenous Randomness from Adversarial Market Learning}
\author{Jian Sun\\FTI Consulting}
\date{\today}

\begin{document}
\maketitle

\begin{abstract}
We propose a deterministic adversarial market model in which apparent randomness emerges endogenously from the interaction between a market mechanism and a population of predictive traders. Unlike a classical generative adversarial network, the model does not attempt to imitate an external empirical data distribution and does not inject random noise into a generator. The market is represented by a deterministic binary return path, while traders learn predictive strategies from observed in-sample history and trade on an out-of-sample continuation. The market then adapts against the traders by reducing their predictive and trading edge.

The central experiment begins with a smooth, highly predictable market path. Traders with multiple lookback windows and multiple holding periods learn to predict future cumulative returns. Initially, these traders earn large out-of-sample profits. After adversarial market adaptation, their out-of-sample profitability collapses toward zero. Importantly, in the final clean specification, no explicit sign-balance, transition-rate, or autocorrelation penalties are imposed. Nevertheless, the out-of-sample return sequence becomes balanced, has transition rate close to one half, has low autocorrelation, and passes block-based distributional diagnostics. In a medium-size experiment with $T_{\mathrm{IS}}=2000$ and $T_{\mathrm{OOS}}=10000$, the out-of-sample positive-return fraction is $0.5010$, the transition rate is $0.4896$, and the maximum absolute autocorrelation is $0.0275$. Binary return blocks transformed into dyadic variables are close to uniform on $[0,1]$, and normalized block sums are broadly consistent with a standard normal law. These results support the hypothesis that market randomness can arise as the endogenous residue of arbitrage pressure rather than from exogenous stochastic shocks.
\end{abstract}

\section{Introduction}

A foundational idea in financial economics is that predictable patterns in market prices should be competed away by traders. In the efficient-market view, the absence of exploitable patterns is not merely a statistical assumption; it is an equilibrium consequence of arbitrage and competition. Standard mathematical finance often models this by postulating an exogenous stochastic process for asset returns, such as Brownian motion, a martingale, or a semimartingale with prescribed distributional properties. The stochasticity is introduced at the beginning.

This paper explores the reverse direction. We ask whether random-looking market behavior can emerge from a deterministic adversarial mechanism without injecting any external random numbers. The model starts with a deterministic and highly predictable market. Traders enter, learn from past returns, and attempt to profit from future returns. The market then adapts to remove the trading edge discovered by those traders. Randomness is therefore not assumed; it is produced as the residual output of a deterministic game between the market and arbitragers.

The construction is inspired by, but distinct from, generative adversarial networks. In a classical GAN, a generator maps exogenous noise into synthetic samples, while a discriminator attempts to distinguish generated samples from real samples. In our setting there is no external data distribution to imitate and no exogenous noise input. The market itself is the object being trained. The trader is not a real/fake discriminator; it is an economic agent that predicts future returns and takes positions. The market objective is not to fool a classifier in the usual statistical sense, but to eliminate the trader's expected trading edge.

The experiments in this paper focus on a binary market model. Each return is constrained to be either $-1$ or $+1$. This binary restriction is intentional: it removes volatility collapse and distributional scale effects, allowing the experiment to focus on predictability. The market begins as a smooth deterministic binary sequence with long predictable regimes. Traders observe rolling windows of past returns and predict the sign of future cumulative returns over different holding periods. We include both fast traders, who predict one-period returns, and slower traders, who predict cumulative returns over two, five, ten, or twenty periods. Thus the trader population can exploit both short-term and medium-term trends.

The main numerical finding is that the out-of-sample market becomes random-looking even when direct statistical regularization is removed. In the final clean model, sign imbalance, transition-rate deviation, and autocorrelation are not penalized. These quantities are recorded only as diagnostics. Nevertheless, after adversarial training, the out-of-sample path has positive-return fraction close to one half, transition rate close to one half, and low maximum autocorrelation. Moreover, block transformations of the binary returns generate variables that are close to uniform, and normalized block sums are close to normal. This suggests that the market learns to remove exploitable structure because traders exploit it, not because the model designer directly imposes randomness.

The rest of the paper is organized as follows. Section~\ref{sec:gan} explains the conceptual difference between the present model and classical GANs. Section~\ref{sec:market} defines the deterministic binary market. Section~\ref{sec:traders} defines the multi-horizon trader population. Section~\ref{sec:training} specifies the market-trader training game. Section~\ref{sec:algorithm} describes the simulation algorithm. Section~\ref{sec:results} reports numerical results. Section~\ref{sec:diagnostics} introduces distributional diagnostics for the final market path. Section~\ref{sec:discussion} discusses interpretation, limitations, and possible extensions.

\section{Relation to GANs and Adversarial Learning}\label{sec:gan}

Classical generative adversarial networks consist of two components: a generator and a discriminator. The generator receives random noise $z$ and maps it to generated samples. The discriminator attempts to distinguish generated samples from real samples. Training is adversarial: the generator improves by learning to fool the discriminator, and the discriminator improves by learning to separate generated from real data.

Our construction is adversarial but not a classical GAN. First, there is no exogenous noise vector. The market is deterministic. Given the initial market path, trader architecture, learning rates, and number of training steps, the final return path is fully determined. Second, there is no external empirical distribution to match. The model is not trained to reproduce historical market data. Instead, it is trained to remove the profitability of predictive trading strategies. Third, the adversary is an economic trader rather than a statistical real/fake discriminator. A trader observes past returns, predicts a future cumulative return, takes a position, and earns profit or loss. The market objective is to make this trading edge vanish.

Therefore, the model is better described as an adversarial self-generated market model. Its purpose is not data imitation but endogenous efficiency formation.

\section{Deterministic Binary Market}\label{sec:market}

Let $T_{\mathrm{IS}}$ and $T_{\mathrm{OOS}}$ denote the lengths of the in-sample and out-of-sample regions. The market is represented by a deterministic vector of latent parameters
\[
\theta = (\theta_1,\ldots,\theta_T),\qquad T=T_{\mathrm{IS}}+T_{\mathrm{OOS}}.
\]
The realized return is binary:
\[
r_t\in\{-1,+1\}.
\]

In the numerical implementation, the hard binary return is obtained from a smooth latent return through a straight-through estimator. Define
\[
\widetilde r_t = \tanh(\beta \theta_t),
\]
where $\beta>0$ is a temperature parameter. The forward return is
\[
r_t=\operatorname{sign}(\widetilde r_t),
\]
with values in $\{-1,+1\}$. During backpropagation, gradients flow through the smooth $\tanh$ approximation. This allows the market to remain binary in the observed path while still being trainable by gradient methods.

The initial latent path is chosen deterministically and smoothly, for example as a low-frequency combination of linear and sinusoidal components:
\[
\theta_t=2(x_t-1/2)+0.8\sin(2\pi x_t)+0.3\sin(4\pi x_t),
\]
where $x_t$ is a normalized time grid on $[0,1]$. This initial path generates long predictable regimes. No random initialization is used.

The market path is divided into an in-sample segment
\[
(r_1,\ldots,r_{T_{\mathrm{IS}}})
\]
and an out-of-sample segment
\[
(r_{T_{\mathrm{IS}}+1},\ldots,r_T).
\]
Traders learn only from the in-sample region. Their performance is evaluated on the out-of-sample region.

\section{Multi-Holding-Period Traders}\label{sec:traders}

A trader is indexed by two parameters:
\[
m=\text{lookback window},\qquad h=\text{holding period}.
\]
The trader observes the recent return window
\[
H_t^{(m)}=(r_t,r_{t+1},\ldots,r_{t+m-1})
\]
and predicts the sign of the next $h$-period cumulative return
\[
S_{t,h}=r_{t+m}+r_{t+m+1}+\cdots+r_{t+m+h-1}.
\]

The trader is a neural network $D_{m,h}$ that outputs a logit. The predicted probability of a positive future cumulative return is
\[
p_t^{(m,h)}=\sigma\!\left(D_{m,h}(H_t^{(m)})\right),
\]
where $\sigma$ is the logistic function.

In the constant-notional version, the trader takes the position
\[
\pi_t^{(m,h)}=\begin{cases}
+1, & p_t^{(m,h)}>1/2,\\
-1, & p_t^{(m,h)}\le 1/2.
\end{cases}
\]
The corresponding realized profit is
\[
\mathrm{PnL}_t^{(m,h)}=\pi_t^{(m,h)}S_{t,h}.
\]

In the variable-notional version, the position is scaled by confidence:
\[
\pi_t^{(m,h)}=2p_t^{(m,h)}-1.
\]
The present paper focuses on the constant-notional version, because it gives a clear measure of directional exploitable structure. The variable-notional version is a natural robustness check.

The default trader population uses
\[
m\in\{3,5,10,20,40\},\qquad h\in\{1,2,5,10,20\}.
\]
This creates traders with different turnover frequencies. A one-period trader exploits next-period predictability; a twenty-period trader exploits medium-term trend predictability.

\section{Market-Trader Training Game}\label{sec:training}

Training proceeds in alternating steps. First, the traders train on the current in-sample market path. For each $(m,h)$, the trader minimizes binary cross-entropy for the label
\[
Y_{t,h}=\mathbf{1}_{\{S_{t,h}>0\}}.
\]
The trader loss is
\[
L_{\mathrm{trader}}^{(m,h)}=\operatorname{BCE}\!\left(D_{m,h}(H_t^{(m)}),Y_{t,h}\right).
\]

Second, the market updates against the trained traders. In the clean final specification, the market loss has only two components. The first component makes traders uncertain in-sample:
\[
L_{\mathrm{IS}}=\frac{1}{|\mathcal{T}|}\sum_{(m,h)\in\mathcal{T}}\operatorname{BCE}\!\left(D_{m,h}(H_t^{(m)}),1/2\right),
\]
where $\mathcal{T}$ is the set of trader types.

The second component removes out-of-sample expected trading edge. For each trader, using a differentiable soft position, we compute expected out-of-sample PnL and penalize its square:
\[
L_{\mathrm{OOS}}=\frac{1}{|\mathcal{T}|}\sum_{(m,h)\in\mathcal{T}}\left(\mathbb{E}_{\mathrm{OOS}}\left[\pi_t^{(m,h)}S_{t,h}\right]\right)^2.
\]
The final market objective is
\[
L_{\mathrm{market}}=\lambda_{\mathrm{IS}}L_{\mathrm{IS}}+\lambda_{\mathrm{OOS}}L_{\mathrm{OOS}}.
\]

No direct penalty is imposed on sign balance, transition rate, or autocorrelation. These are diagnostics only. This is the key modeling choice. If the final return path becomes balanced or low-autocorrelation, it does so because such patterns were exploitable by traders, not because they were imposed by the loss function.

\section{Algorithm}\label{sec:algorithm}

\begin{algorithm}[h!]
\caption{Adversarial deterministic market learning}\label{alg:market}
\begin{algorithmic}[1]
\STATE Initialize a deterministic smooth binary market path.
\STATE Split the path into in-sample and out-of-sample regions.
\STATE Initialize a family of traders $D_{m,h}$ deterministically.
\FOR{each outer epoch}
    \STATE Train all traders on the in-sample region.
    \STATE Evaluate trader PnL on the out-of-sample region.
    \STATE Update the market by minimizing $L_{\mathrm{market}}$.
    \STATE Record diagnostics: trader accuracy, trader PnL, sign fraction, transition rate, autocorrelation, and distributional tests.
\ENDFOR
\STATE At the final epoch, perform distributional diagnostics on the in-sample and out-of-sample paths.
\end{algorithmic}
\end{algorithm}

\section{Numerical Design}\label{sec:numerical_design}

The main experiment uses the clean constant-notional specification. The market return is binary, the initial path is deterministic and smooth, and the trader population contains $25$ traders formed by the Cartesian product of five lookback windows and five holding periods. The clean objective uses only the in-sample trader-uncertainty term and the out-of-sample expected-edge term.

The medium-size experiment reported below uses
\[
T_{\mathrm{IS}}=2000,\qquad T_{\mathrm{OOS}}=10000,
\]
with $150$ outer epochs in the numerical code. The baseline experiment with $T_{\mathrm{IS}}=1000$ and $T_{\mathrm{OOS}}=3000$ gives qualitatively similar results, but the medium experiment gives more stable distributional diagnostics because it produces more non-overlapping blocks for the uniform and normal quantile tests.

\begin{table}[h!]
\centering
\caption{Main numerical configuration.}
\label{tab:numerical_config}
\begin{tabular}{lc}
\toprule
Item & Value \\
\midrule
In-sample length $T_{\mathrm{IS}}$ & $2000$ \\
Out-of-sample length $T_{\mathrm{OOS}}$ & $10000$ \\
Lookback windows $m$ & $\{3,5,10,20,40\}$ \\
Holding periods $h$ & $\{1,2,5,10,20\}$ \\
Number of trader types & $25$ \\
Return values & $\{-1,+1\}$ \\
Direct sign-balance penalty & none \\
Direct transition-rate penalty & none \\
Direct autocorrelation penalty & none \\
\bottomrule
\end{tabular}
\end{table}

\section{Numerical Results}\label{sec:results}

The initial market is highly predictable. In the baseline experiment, before market adaptation, the out-of-sample trader accuracy is approximately $0.992$, and the average out-of-sample trader PnL is about $22016.6$. The out-of-sample transition rate is approximately $0.001$, and maximum absolute autocorrelation is approximately $0.997$. This confirms that the initial market is almost deterministic and highly exploitable.

After adversarial training, the out-of-sample market becomes difficult to exploit. In the medium clean constant-notional experiment, the final out-of-sample average accuracy is approximately $0.5857$, while the final average out-of-sample PnL is approximately $-79.44$. The positive-return fraction is approximately $0.5010$, the transition rate is approximately $0.4896$, and the maximum absolute autocorrelation is approximately $0.0275$.

The average classification accuracy can remain above $1/2$ even when economic PnL is close to zero, particularly for multi-period holding periods where the sign label is affected by zero or imbalanced cumulative returns. For this reason, the economically relevant diagnostic is out-of-sample PnL rather than raw classification accuracy. The final PnL is small relative to the initial exploitable PnL scale.

These statistics are notable because no direct sign-balance, transition-rate, or autocorrelation penalty was used. The market becomes statistically balanced and low-autocorrelation as a consequence of the trader-market game.

\begin{table}[h!]
\centering
\caption{Selected final diagnostics for the medium clean constant-notional experiment.}
\label{tab:oos_summary}
\begin{tabular}{lcc}
\toprule
Diagnostic & IS & OOS \\
\midrule
Average trader accuracy & $0.7356$ & $0.5857$ \\
Average trader PnL & $2154.48$ & $-79.44$ \\
Positive-return fraction & $0.4320$ & $0.5010$ \\
Transition rate & $0.4922$ & $0.4896$ \\
Maximum absolute autocorrelation & $0.0806$ & $0.0275$ \\
\bottomrule
\end{tabular}
\end{table}

\begin{table}[h!]
\centering
\caption{Final out-of-sample performance by holding period.}
\label{tab:oos_by_horizon}
\begin{tabular}{ccc}
\toprule
Holding period $h$ & Average accuracy & Average final PnL \\
\midrule
$1$  & $0.5008$ & $16.8$ \\
$2$  & $0.7346$ & $-50.0$ \\
$5$  & $0.4974$ & $-2.4$ \\
$10$ & $0.6117$ & $-146.0$ \\
$20$ & $0.5839$ & $-215.6$ \\
\bottomrule
\end{tabular}
\end{table}

\begin{figure}[h!]
\centering
\includegraphics[width=0.95\textwidth]{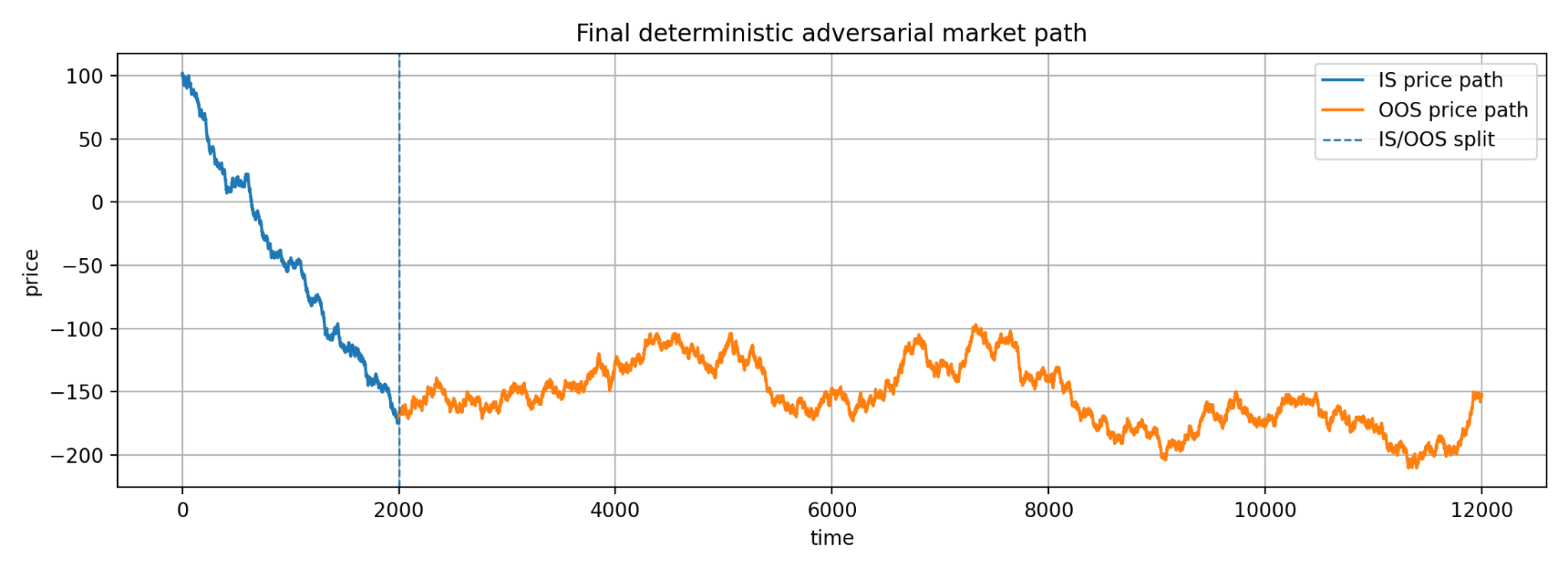}
\caption{Final market path. The figure shows the in-sample and out-of-sample regions after adversarial training. The out-of-sample continuation is no longer the smooth predictable path present at initialization.}
\label{fig:final_market_path}
\end{figure}

\begin{figure}[h!]
\centering
\includegraphics[width=0.95\textwidth]{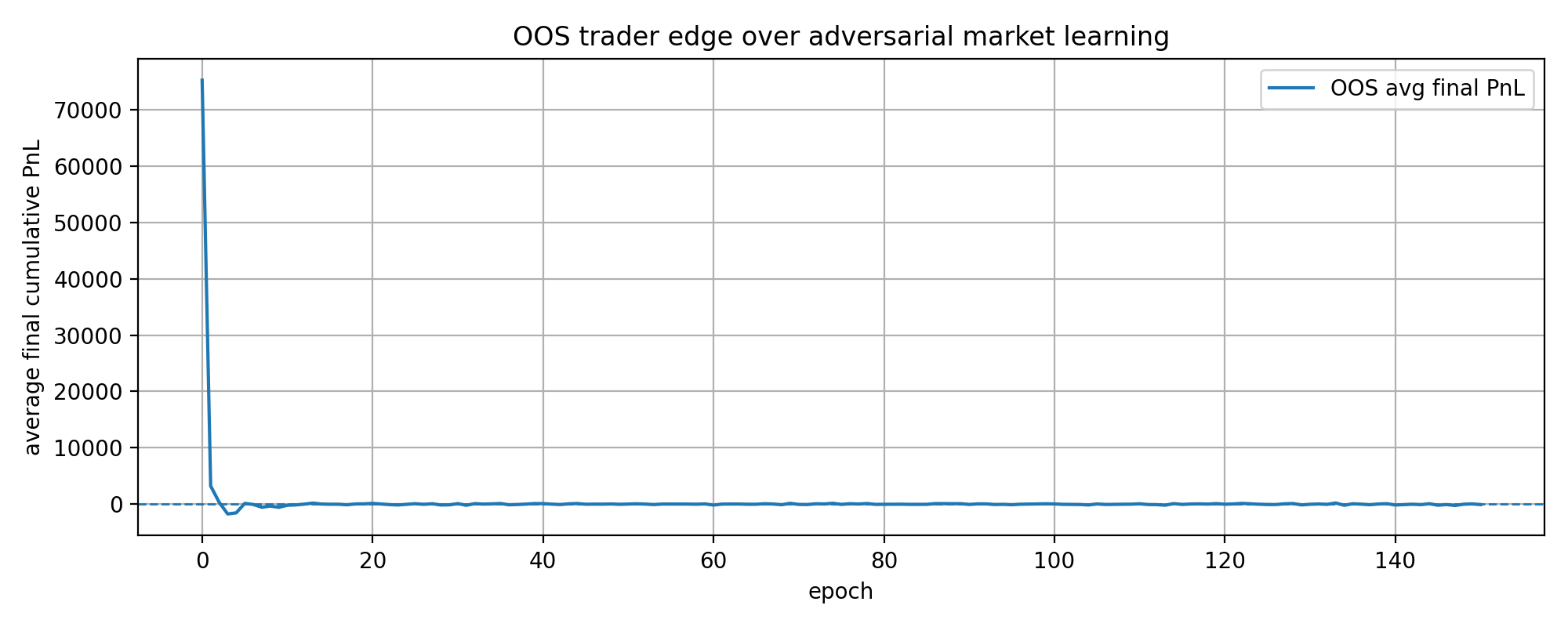}
\caption{Collapse of out-of-sample trader PnL during adversarial training. This figure illustrates the main economic mechanism: predictable trading edge is gradually competed away by market adaptation.}
\label{fig:oos_pnl_collapse}
\end{figure}

\begin{figure}[h!]
\centering
\includegraphics[width=0.95\textwidth]{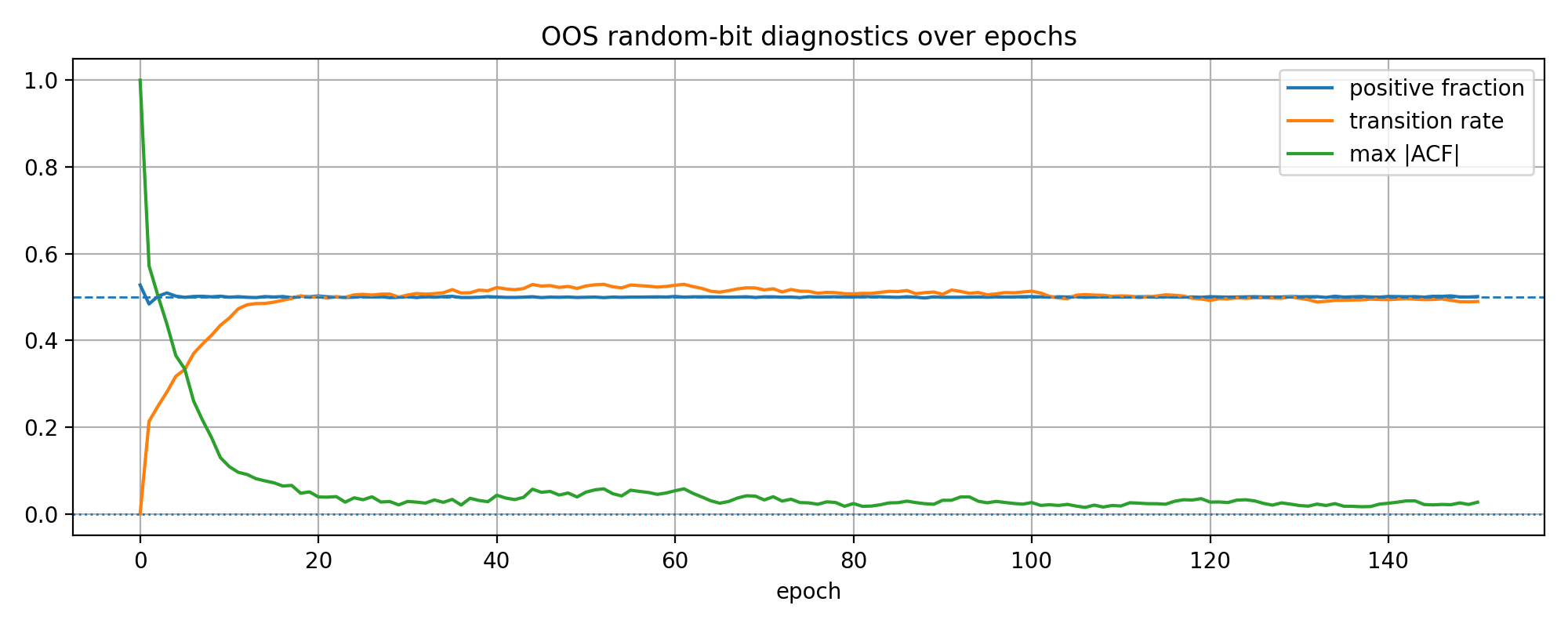}
\caption{Out-of-sample random-bit diagnostics across training epochs. The positive-return fraction and transition rate move toward one half, while the maximum absolute autocorrelation falls, even though none of these quantities is directly penalized in the clean objective.}
\label{fig:oos_random_bit_diagnostics}
\end{figure}

\FloatBarrier

\section{Distributional Diagnostics}\label{sec:diagnostics}

Because the market returns are binary, we introduce three diagnostics to study whether the final sequence behaves like random bits.

\subsection{Trader Probability Distribution}

For each trader, we examine
\[
p_t^{(m,h)}=D_{m,h}(H_t^{(m)}).
\]
If traders are uncertain, these probabilities should concentrate near $1/2$. In the reported medium experiment, probabilities are not always concentrated near $1/2$. The out-of-sample pooled probabilities have mean about $0.2959$ and median about $0.2570$. This indicates that probability uncertainty is not the only mechanism for eliminating edge. Traders can remain directionally biased or confident while their predictions are not aligned with profitable realized outcomes.

This diagnostic should therefore be interpreted carefully. The primary economic target is not $p_t\equiv 1/2$ pointwise, but rather vanishing expected trading edge:
\[
\mathbb{E}_{\mathrm{OOS}}\left[\pi_t^{(m,h)}S_{t,h}\right]\approx 0.
\]

\subsection{Block-to-Uniform Transformation}

Convert binary returns into bits:
\[
b_t=\mathbf{1}_{\{r_t=+1\}}.
\]
For a block size $K$, define
\[
U_j=\sum_{i=1}^{K} b_{jK+i}2^{-i}.
\]
If the bits behave like fair coin flips, $U_j$ should be close to uniform on $[0,1]$.

The out-of-sample block-to-uniform diagnostics are strong in the medium experiment. For block size $K=8$, the mean is approximately $0.4997$, the standard deviation is approximately $0.2929$, and the Kolmogorov--Smirnov statistic versus uniform is approximately $0.0171$. For $K=10$, the mean is approximately $0.4959$, the standard deviation is approximately $0.2877$, and the Kolmogorov--Smirnov statistic is approximately $0.0211$. These values are close to the uniform benchmark, for which the standard deviation is $1/\sqrt{12}\approx 0.2887$.

\begin{table}[h!]
\centering
\caption{Out-of-sample block-to-uniform diagnostics.}
\label{tab:uniform_diagnostics}
\begin{tabular}{ccccc}
\toprule
Block size $K$ & Count & Mean & Std. dev. & KS statistic \\
\midrule
$8$  & $1250$ & $0.4997$ & $0.2929$ & $0.0171$ \\
$10$ & $1000$ & $0.4959$ & $0.2877$ & $0.0211$ \\
$16$ & $625$  & $0.4892$ & $0.2938$ & $0.0319$ \\
\bottomrule
\end{tabular}
\end{table}

\begin{figure}[h!]
\centering
\includegraphics[width=0.75\textwidth]{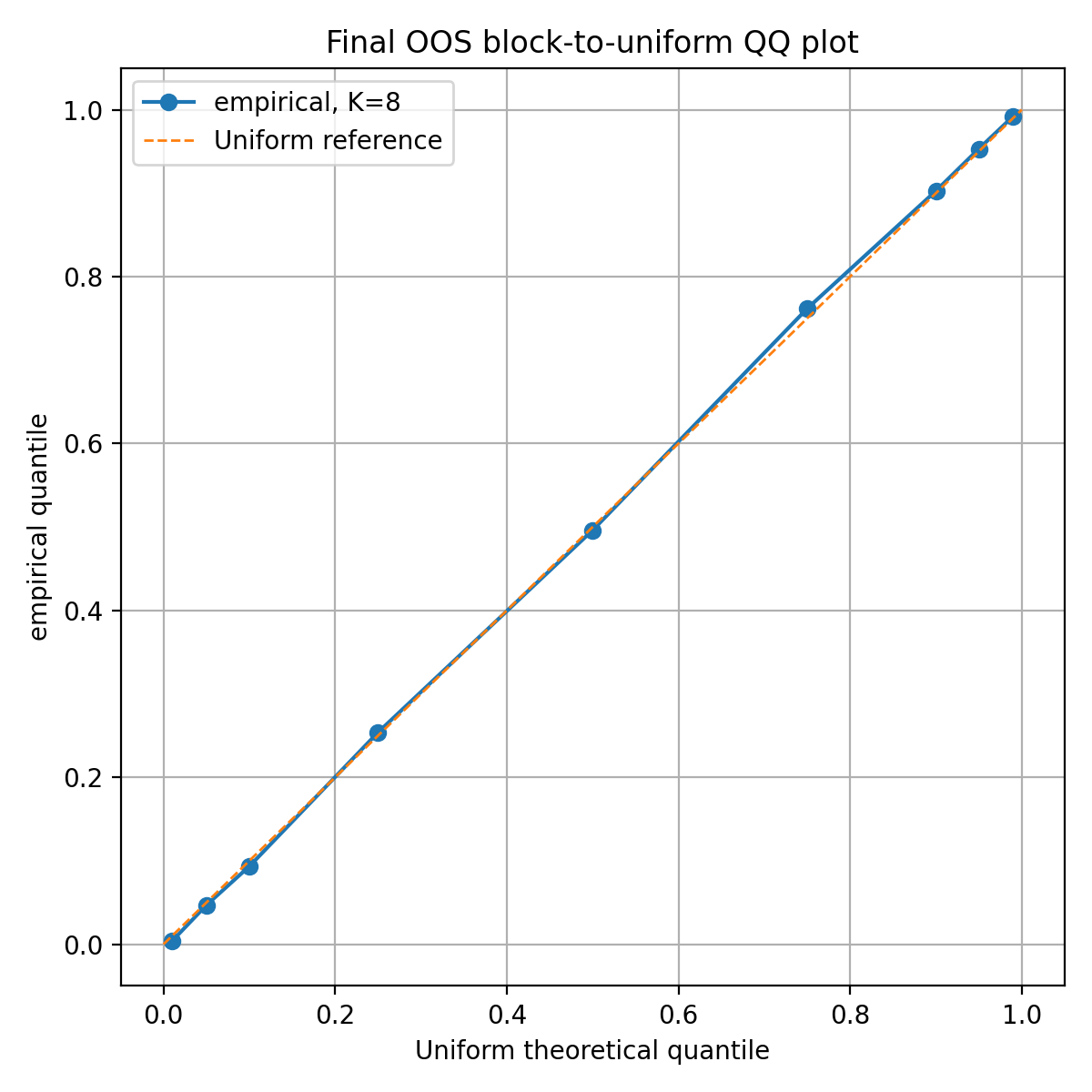}
\caption{Out-of-sample uniform quantile plot for the block-to-uniform transformation. The proximity to the $45^\circ$ line indicates that the final deterministic binary sequence behaves like random bits under this block transformation.}
\label{fig:uniform_qq}
\end{figure}

Thus, the final out-of-sample binary sequence produces nearly uniform block variables, despite being generated by a deterministic adversarial mechanism with no random input.

\subsection{Normalized Block Sums}

For a block size $K$, define
\[
Z_j=\frac{1}{\sqrt{K}}\sum_{i=1}^{K}r_{jK+i}.
\]
If the returns behave like independent fair binary increments, $Z_j$ should be approximately standard normal for moderately large $K$.

For the out-of-sample path, normalized block sums are broadly consistent with this behavior. For $K=10$, the mean is approximately $0.0063$ and the standard deviation is approximately $1.0064$. For $K=20$, the mean is approximately $0.0089$ and the standard deviation is approximately $1.0465$. For $K=100$, the mean is approximately $0.0200$ and the standard deviation is approximately $1.0436$. Quantiles are discrete because the underlying returns are binary, but the central part of the distribution aligns reasonably with the standard normal benchmark.

\begin{table}[h!]
\centering
\caption{Out-of-sample normalized block-sum diagnostics.}
\label{tab:normal_diagnostics}
\begin{tabular}{ccccc}
\toprule
Block size $K$ & Count & Mean & Std. dev. & KS statistic \\
\midrule
$10$  & $1000$ & $0.0063$ & $1.0064$ & $0.1290$ \\
$20$  & $500$  & $0.0089$ & $1.0465$ & $0.1060$ \\
$50$  & $200$  & $0.0141$ & $1.0836$ & $0.0850$ \\
$100$ & $100$  & $0.0200$ & $1.0436$ & $0.0693$ \\
\bottomrule
\end{tabular}
\end{table}

\begin{figure}[h!]
\centering
\includegraphics[width=0.75\textwidth]{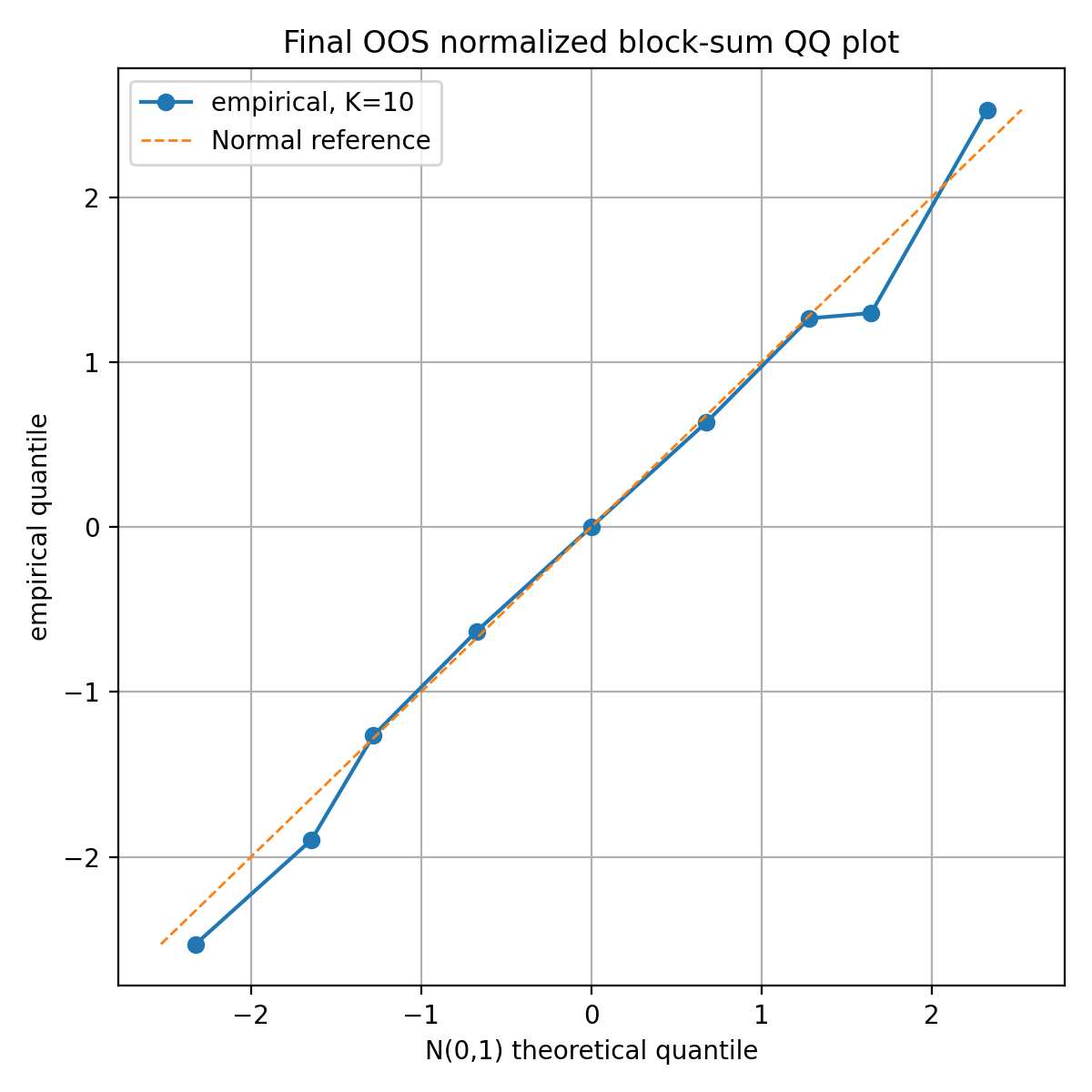}
\caption{Out-of-sample normal quantile plot for normalized block sums. Discreteness remains visible because returns are binary, but the final block sums are centered near zero and have approximately unit scale.}
\label{fig:normal_qq}
\end{figure}

\begin{figure}[h!]
\centering
\includegraphics[width=0.75\textwidth]{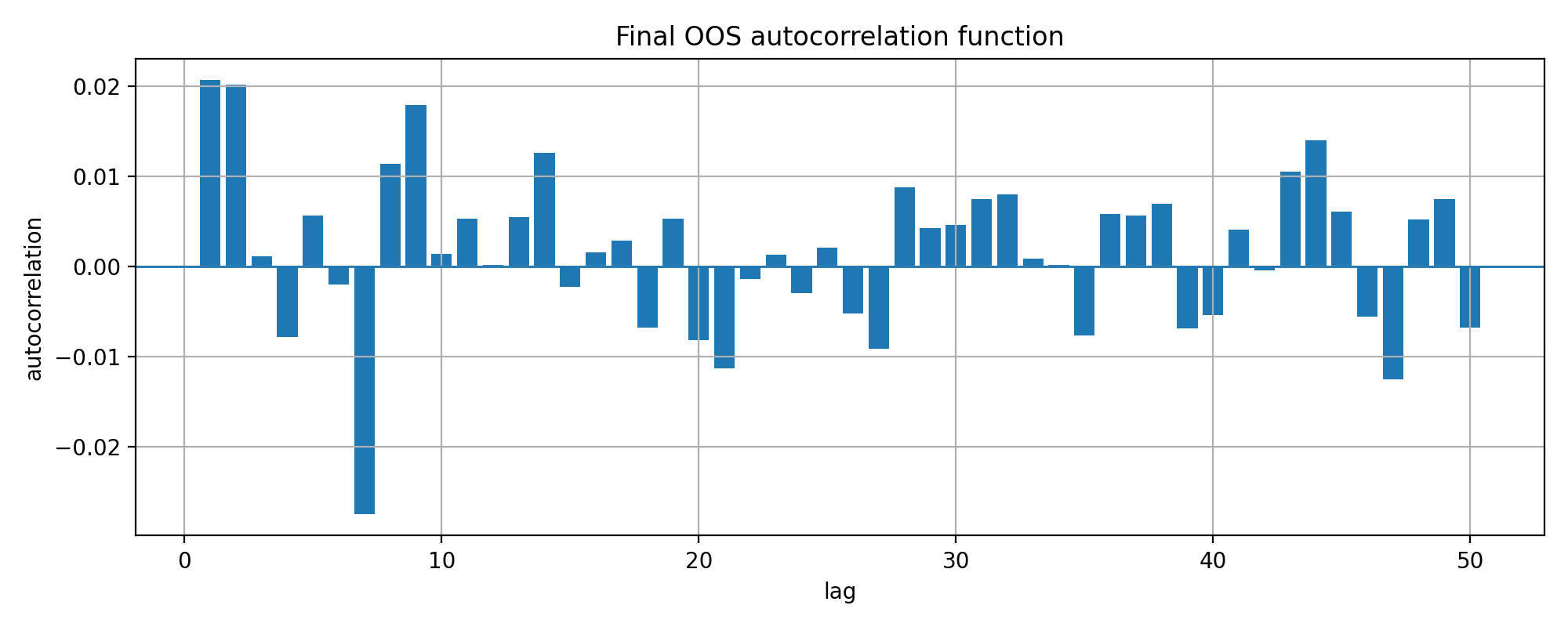}
\caption{Final out-of-sample autocorrelation function. The maximum absolute autocorrelation is small, despite the absence of a direct autocorrelation penalty in the clean objective.}
\label{fig:oos_acf}
\end{figure}

\FloatBarrier

\section{Interpretation}\label{sec:discussion}

The experiments support the following interpretation. A market does not need exogenous stochastic shocks to become random-looking. If traders are allowed to exploit predictable structure across several horizons, the market can adapt endogenously until the remaining sequence appears statistically random relative to those traders.

This provides a constructive mechanism for market efficiency. Instead of assuming randomness as a primitive, the model derives random-looking behavior as an equilibrium residue of adversarial learning.

The out-of-sample results are especially important. The in-sample path remains structured and predictable, but traders trained on that structure lose their out-of-sample edge. This resembles the empirical phenomenon that strategies discovered from historical data often decay after they become known or crowded.

The distributional diagnostics strengthen this interpretation. The block-to-uniform variables are close to uniform, and the normalized block sums are close to standard normal. These are not imposed distributional targets. They arise because the trader population removes exploitable patterns from the future return sequence.

\section{Limitations and Extensions}

The present model is intentionally simple. The market return is binary, and the trader networks are small feed-forward predictors. The market path is finite and deterministic. There is no transaction cost, no risk constraint, no capital constraint, no heterogeneous objective function, and no market impact model beyond the adversarial update itself.

Several extensions are natural. First, the trader population can be enriched. In addition to neural-network traders, one could introduce explicit trend-following traders, mean-reversion traders, run-length traders, frequency-domain traders, and volatility-regime traders. Second, position sizing can be made variable. Instead of fixed notional positions, the trader can trade
\[
\pi_t=2p_t-1,
\]
or use a risk-adjusted position based on predicted edge and volatility. Third, one can move from binary returns to continuous capped returns. This requires a nonzero activity constraint to prevent the trivial zero-return solution. Fourth, one can study strict out-of-sample experiments in which the market adapts only on in-sample information and the out-of-sample region is reserved solely for testing. Finally, one can develop theoretical conditions under which adversarial removal of a sufficiently rich family of trading strategies implies martingale-like or random-walk-like properties.

\section{Conclusion}

We have presented a deterministic adversarial market model in which random-looking returns emerge endogenously from the interaction between a market and a population of predictive traders. The market begins smooth and predictable. Traders learn profitable patterns. The market then adapts to remove their edge.

In the clean binary multi-holding-period experiment, no random input is used and no direct statistical penalties are imposed on sign balance, transition rate, or autocorrelation. Nevertheless, the out-of-sample market becomes balanced, low-autocorrelation, and close to random-bit behavior under block-to-uniform and normalized block-sum diagnostics.

The results suggest that market randomness can be viewed not merely as an exogenous modeling assumption, but as the endogenous outcome of arbitrage pressure. Randomness, in this perspective, is the residual structure left after a sufficiently rich class of traders has competed away predictable profit opportunities.

\bibliographystyle{plainnat}

\end{document}